\documentclass[11pt]{article}
\usepackage{moriond,graphicx}

\newcommand{\epem}   {\ensuremath{\mathrm{e^+e^-}}}

\newcommand{\oaa}    {\ensuremath{\mathcal{O}(\alpha_s^2)}}
\newcommand{\bt}     {\ensuremath{B_{\mathrm{T}}}}
\newcommand{\bw}     {\ensuremath{B_{\mathrm{W}}}}
\newcommand{\mh}     {\ensuremath{M_{\mathrm{H}}}}
\newcommand{\mhsq}   {\ensuremath{M_{\mathrm{H}}^2}}
\newcommand{\thr}    {\ensuremath{1-T}}
\newcommand{\cp}     {\ensuremath{C}}
\newcommand{\ca}     {\ensuremath{C_{\mathrm{A}}}}
\newcommand{\cf}     {\ensuremath{C_{\mathrm{F}}}}
\newcommand{\tf}     {\ensuremath{T_{\mathrm{F}}}}
\newcommand{\nf}     {\ensuremath{n_{\mathrm{f}}}}
\newcommand{\anull}  {\ensuremath{\alpha_0}}
\newcommand{\anulltwo}{\ensuremath{\anull(2\;\mathrm{GeV})}}

\newcommand{\as}     {\ensuremath{\alpha_{\mathrm{S}}}}

\newcommand{\mui}    {\ensuremath{\mu_{\mathrm{I}}}}

\newcommand{\roots}  {\ensuremath{\sqrt{s}}}

\newcommand{\asmz}   {\ensuremath{\as(M_{\mathrm{Z^0}})}}
\newcommand{\sigl}   {\ensuremath{\sigma_{\mathrm{L}}}}

\newcommand{\siglch} {\ensuremath{\sigma_{\mathrm{L}}^{\mathrm{ch}}}}
\newcommand{\sigtch} {\ensuremath{\sigma_{\mathrm{T}}^{\mathrm{ch}}}}

\newcommand{\qcost}  {\ensuremath{q\cdot\cos\Theta}}

\newcommand{\lmsb}   {\ensuremath{\Lambda_{\mathrm{\overline{MS}}}}}
\newcommand{\perc}   {\%}
\newcommand{\rpt}    {\ensuremath{R_{\mathrm{PT}}}}

\begin{document}
\vspace*{4cm}
\title{ EXPERIMENTAL STUDIES OF POWER CORRECTIONS }

\author{ S. KLUTH }

\address{ Max-Planck-Institut f\"ur Physik, F\"ohringer Ring 6,
D-80805 M\"unchen, Germany, \\ skluth@mppmu.mpg.de }

\maketitle

\abstracts{ We present recent results from experimental studies of
power corrections in the model of Dokshitzer, Marchesini and Webber
using data from the process $\epem\rightarrow$ hadrons at several
centre-of-mass energies. Fitted \oaa+NLLA QCD predictions combined
with power corrections to model hadronisation effects successfully
describe differential distributions and mean values of event shape
observables. The fit results for the strong coupling constant and the
free parameter of the power correction calculations from the various
observables are consistent with each other within the
uncertainties. The same fits are further employed to study the gauge
structure of QCD by additionally varying the QCD colour factors in the
fits. The results are consistent with QCD based on the SU(3) symmetry
group. A new measurement of the transverse cross section is shown
using reanalysed data of the former JADE experiment at DESY and its
implications for power correction are discussed.  }

\section{ Tests of Power Corrections }
\label{sec_powcor}

In the process of hadron production in \epem\ annihilation confinement
effects, also referred to as hadronisation, play an important
r\^ole. These effects stemming from the transition between the partons
produced in the \epem\ annihilation and the hadrons observed in the
detector are not directly accessible by perturbative QCD
calculations. Several QCD based hadronisation models implemented in
Monte Carlo simulation programs exist and generally describe the data 
well~\cite{jetset3,herwig,ariadne3}. 

Recently new analytic models of hadronisation have been developed. In
these the structure of non-perturbative contributions surpressed by
powers of the hard scale (power corrections) is deduced from
ambiguities in the perturbative
result~\cite{dokshitzer95a,beneke95a}. In the model of Dokshitzer,
Marchesini and Webber (DMW) the effects of gluon radiation with
transverse momenta ${\cal O}(\lmsb)$, i.e. ${\cal O}(100)$~MeV, are
studied. The model assumes that the physical strong coupling remains
finite in the region around the Landau pole where simple perturbative
evolution of the strong coupling breaks down~\footnote{See the
contribution by D.V. Shirkov to this conference.}. This leads to the
introduction of a free parameter $\anull=1/\mui\int_0^{\mui}\as(k)dk$
which must be determined from the data.

The main prediction of the DMW model is that distributions of the
event shape observables \thr, \mh\ or \mhsq, \bt, \bw\ and \cp\ 
are described by the perturbative QCD prediction \rpt\ shifted by
an amount proportional to $\anull/Q$ where $Q=\roots$ is the hard
scale of the process~\cite{dokshitzer98a,dokshitzer98b}. The free
parameter \anull\ is universal, since obervable specific details are
calculated with a theoretical uncertainty of about 20\perc\
\cite{dokshitzer98b}. For mean values one obtaines an additive
correction to the perturbative prediction also proportional to
$\anull/Q$. The perturbative QCD predictions are \oaa+NLLA in the case
of event shape distributions and \oaa\ in the case of mean values. 

\begin{figure}[!htb]
\begin{center}
\begin{tabular}{cc}
\includegraphics[width=0.475\textwidth]{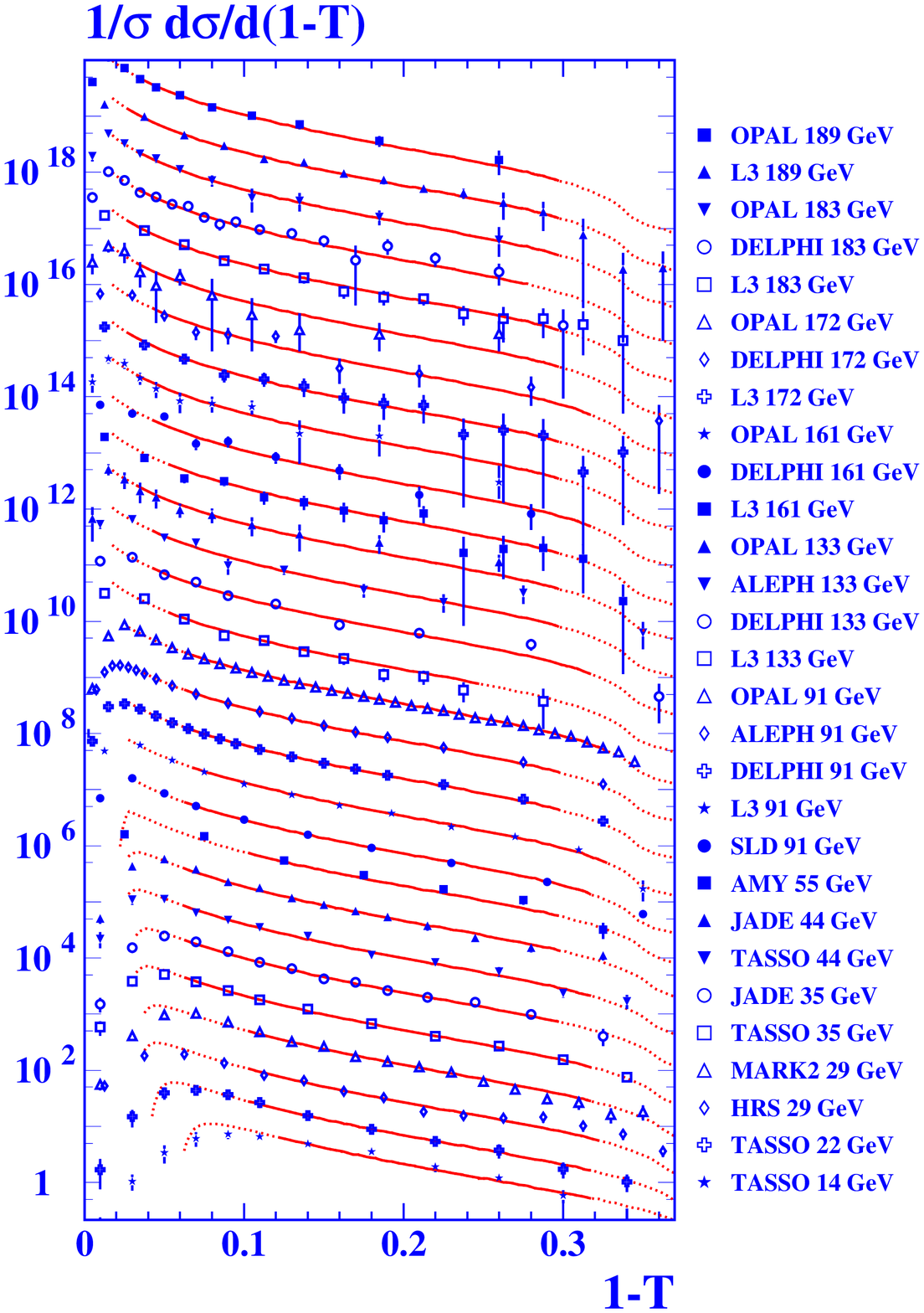} & 
\raisebox{5cm}{
\begin{tabular}{c}
\includegraphics[width=0.325\textwidth]{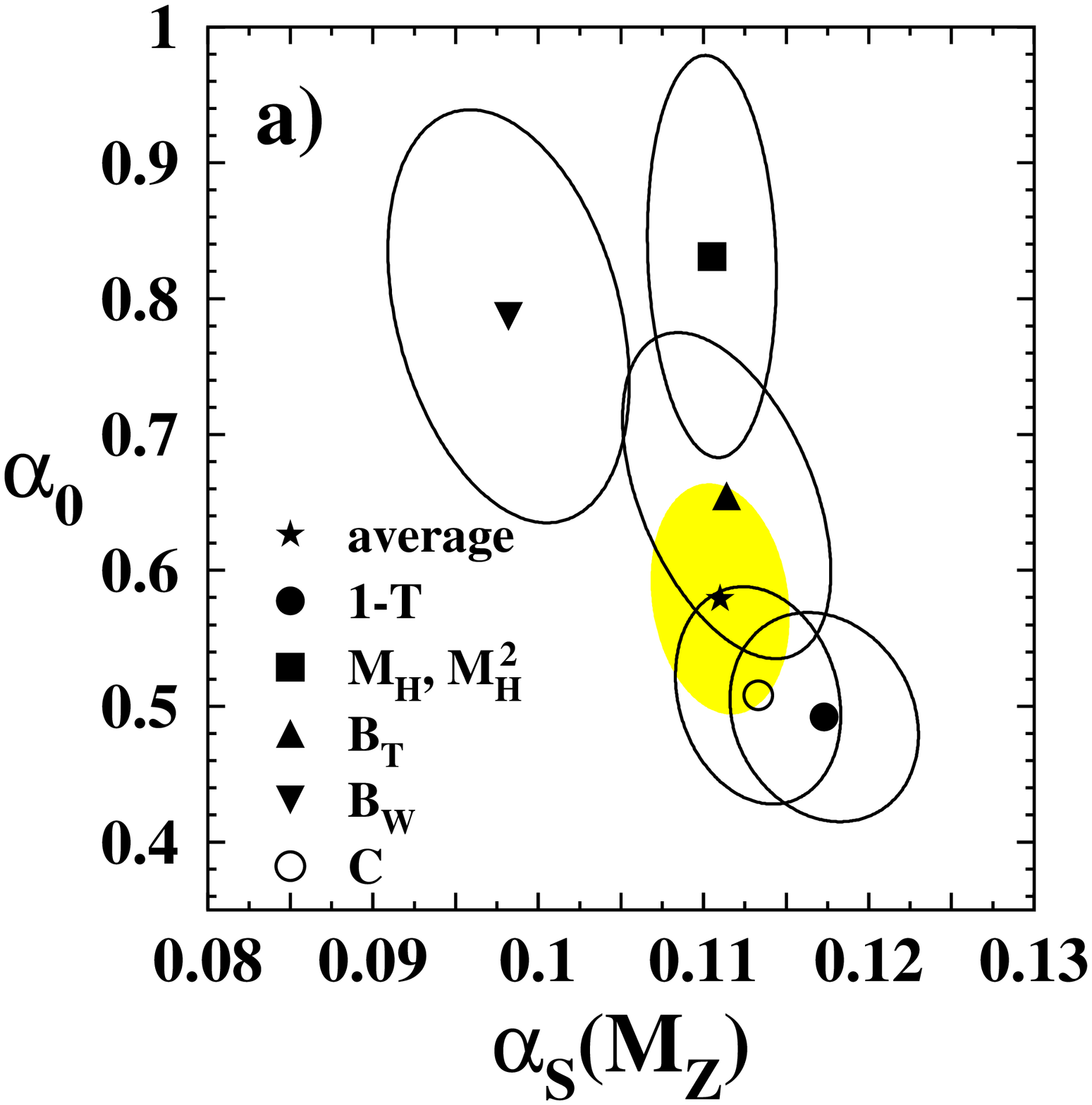} \\
\includegraphics[width=0.325\textwidth]{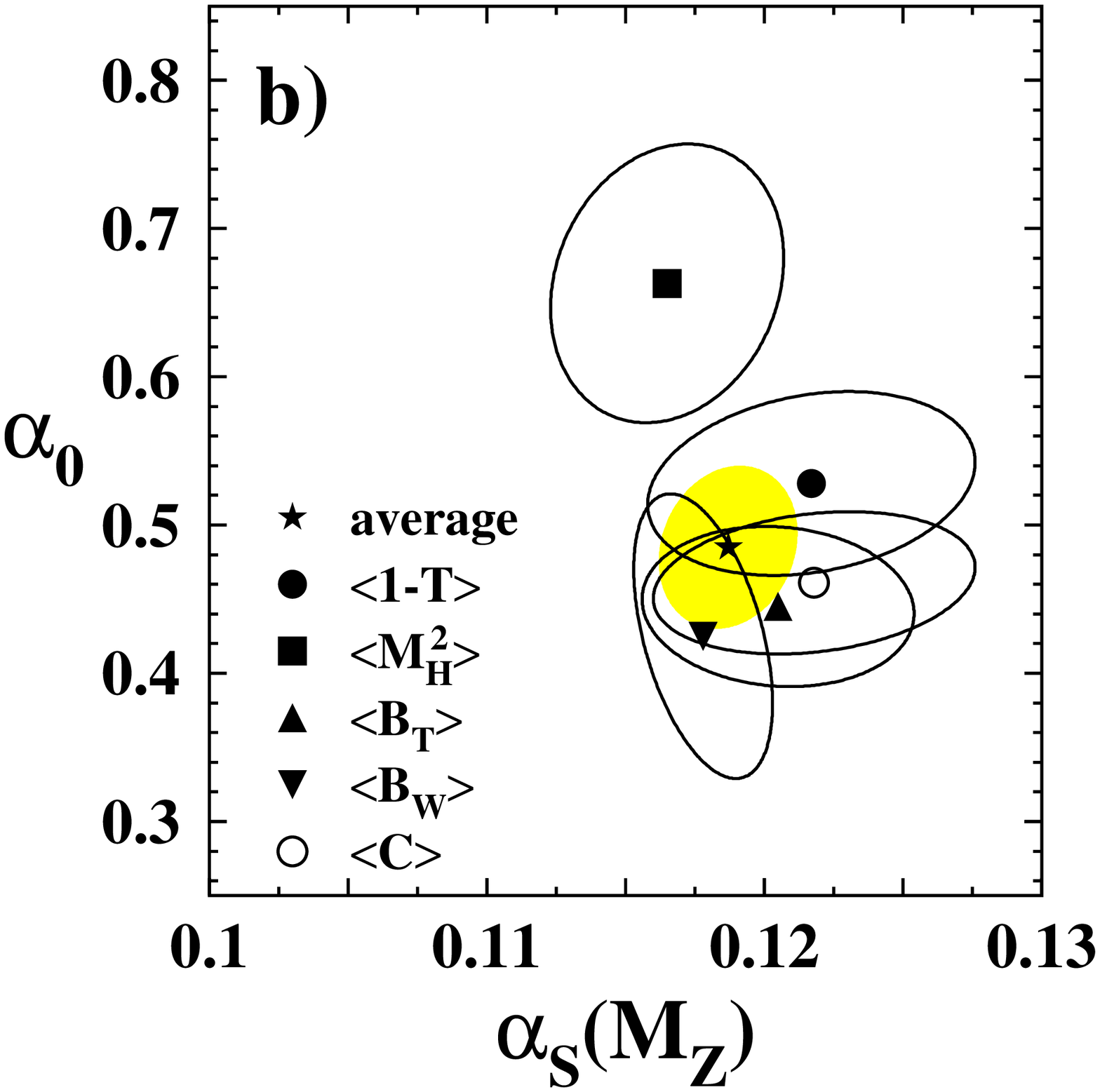} \\
\end{tabular} } \\
\end{tabular}
\caption[bla]{ The figure on the left shows scaled distributions of
\thr\ measured at $\roots=14$ to 189~GeV~\cite{powcor}. The solid
lines represent the fit result while the dotted lines indicate an
extrapolation of the fit. The figures on the right show results for
\asmz\ and \anulltwo\ of fits to distributions (top) and mean values
(bottom) as one standard deviation error ellipses~\cite{powcor}. }
\label{fig_powcor}
\end{center}
\end{figure}

Figure~\ref{fig_powcor} (left) presents the results
of fits with power corrections as described above to distributions of
\thr\ \cite{powcor}. The fit describes the data well at all cms
energies ranging from 14 to 189~GeV. The figures~\ref{fig_powcor}
(right) present the results for \asmz\ and \anulltwo\ from fits to
distributions and mean values. There is reasonable agreement between
the individual results within the total uncertainties. However, the
results for \anulltwo\ from \mh\ or \mhsq\ and from \bw\ for
distributions are large compared to results from the other
observables. This observation may be related to the non-inclusiveness
of these observables, i.e. that only particles contained in one
hemisphere of the event contribute~\footnote{See the contribution by
S. Tafat to this conference.}. 

\section{ Colour Factors }

\begin{figure}[!htb]
\begin{center}
\begin{tabular}{cc}
\includegraphics[width=0.425\textwidth]{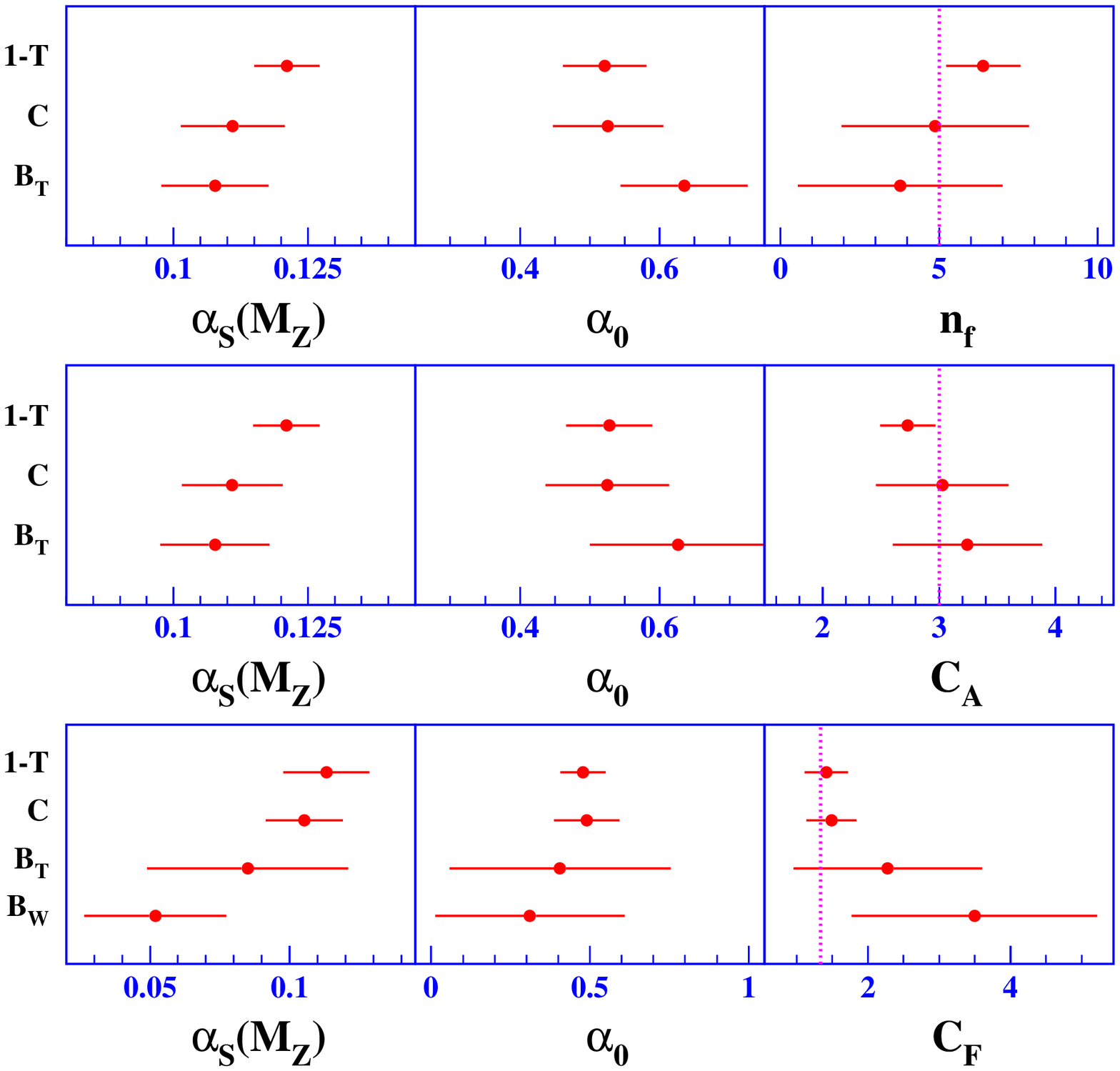} & 
\includegraphics[width=0.425\textwidth]{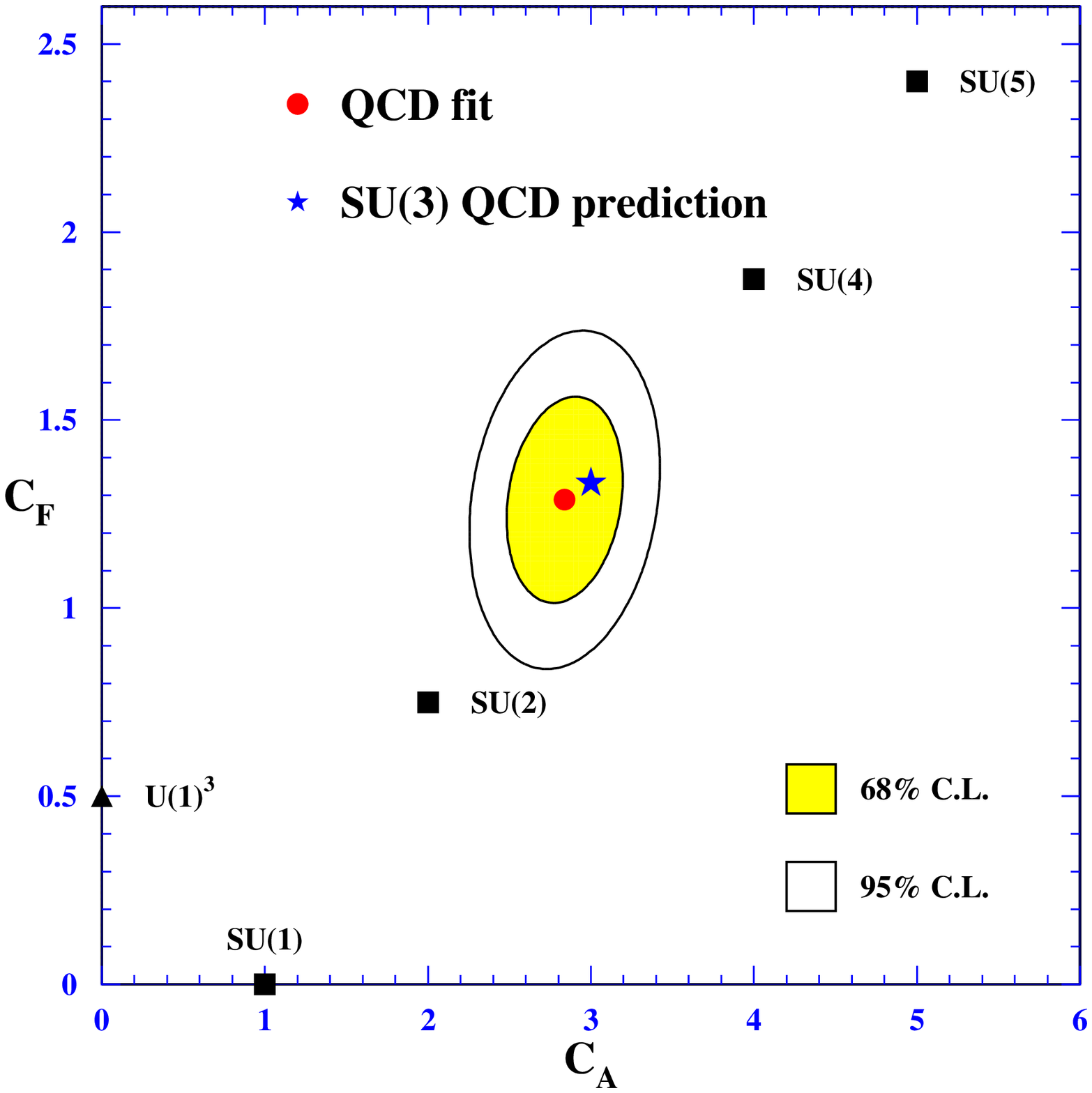} \\
\end{tabular}
\caption[bla]{ The figure on the left shows fit results for \asmz,
\anulltwo\ and one of the colour factors \nf, \ca\ or
\cf~\cite{colrun}. The vertical dotted lines indicate the expectation
from standard QCD for the colour factor. The figure on the right
presents combined results for \ca\ and \cf\ from fits of \asmz, \ca\
and \cf~\cite{colrun}. The square and triangle symbols indicate
expecations for \ca\ and \cf\ for different symmetry groups. }
\label{fig_colrun}
\end{center}
\end{figure}

The same fits as described in section~\ref{sec_powcor} are generalised
to study the gauge structure of QCD by varying the QCD colour factors
$\ca=3$, $\cf=4/3$ or $\tf=1/2$ (or equivalently the number of active
quark flavours $\nf=5$) with the SU(3) symmetry
group~\cite{colrun}. These fits avoid a potential bias of traditional
analyses of the QCD gauge structure using angular correlations in
4-jet final states where hadronisation effects are corrected with
Monte Carlo models based on standard QCD. Figure~\ref{fig_colrun}
(left) shows the results of fits with
\asmz, \anulltwo\ and one of the colour factors as free
parameters. Good agreement with the expectations from standard QCD is
observed. Alternative fits where \anulltwo\ is fixed at a value
determined from previous measurements yield consistent
results. Figure~\ref{fig_colrun} (right) presents the combined results
for \ca\ and \cf\ from simultaneous fits of \asmz, \ca\ and \cf\ to
\thr\ and \cp. The results of the simultaneous fits are also 
consistent with standard QCD while some other possibilities for the
gauge symmetry group of QCD are excluded. Assuming SU(3) to be the
correct gauge symmetry of QCD these results can be interpreted as a
successful consistency check of the DMW model.

\section{ Transverse Cross Section }

\begin{figure}[!htb]
\begin{center}
\begin{tabular}{cc}
\includegraphics[width=0.425\textwidth]{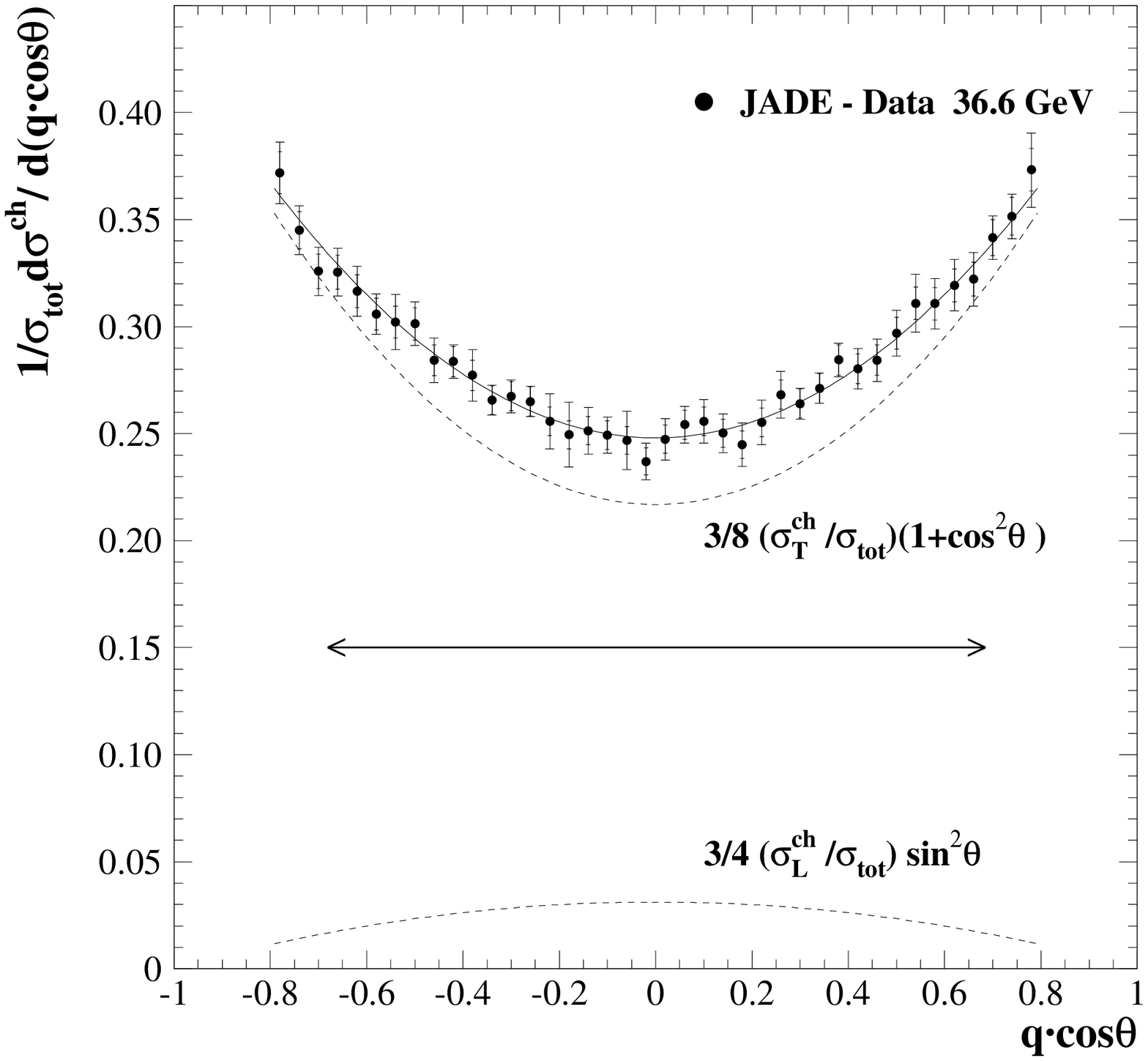} & 
\includegraphics[width=0.425\textwidth]{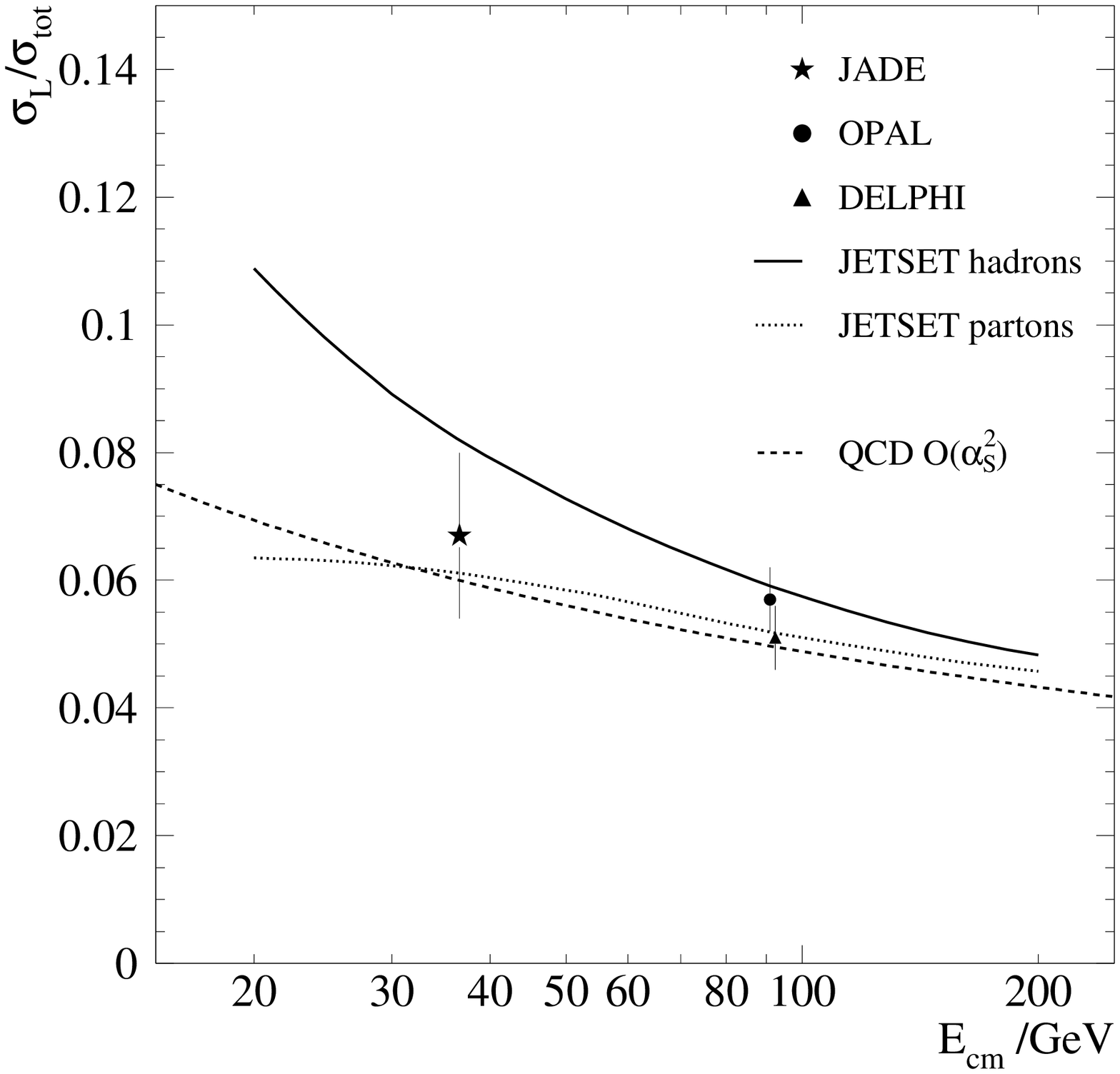} \\
\end{tabular}
\caption[bla]{ The figure on the left shows the distribution of angles
\qcost\ between charged hadrons and the incoming
$e^-$~\cite{sigljade}. Superimposed  
are the fit result (solid line) and its two components proportional to
\sigtch\ and \siglch\ (dashed lines). The figure on the right shows
the result for \sigl\ compared with other measurements and predictions
from the JETSET Monte Carlo simulation and from perturbative
QCD~\cite{sigljade}. } 
\label{fig_sigl}
\end{center}
\end{figure}

A new measurement of longitudinal cross section \sigl\ at
$\roots=36.6$ has been performed~\cite{sigljade} using data of the
former JADE experiment at the PETRA \epem\ collider at
DESY~\cite{naroska87}. The distribution of angles $\Theta$ between the
incoming $e^-$ and the outgoing charged hadrons shown in
figure~\ref{fig_sigl} (left) is sensitive to \sigl. The contribution
$\sim\sigtch(1+\cos^2\Theta)$ is from the elektroweak interaction with
transverse polarisation while the contribution
$\siglch\sim\sin^2\Theta$ stems from gluon radiation of the primary
quarks. The analysis finds $\sigl=0.067\pm0.013$ which is translated
in \oaa\ into a measurement of the strong coupling constant
$\as(36.6\;\mathrm{GeV})=0.150\pm0.025$. Figure~\ref{fig_sigl} (right)
shows the measured $\sigl(36.6\;\mathrm{GeV})$ together with other
measurements and compared to expectations from the JETSET Monte Carlo
and from perturbative QCD. The predictions are in good agreement with
the measurements within the errors.

The power correction to \sigl\ is expected as
$\delta_{\mathrm{PC}}\sim\mui/Q(\anull(\mui)-\as)$
\cite{dokshitzer95}. Fitting the full expression for the power
correction to the data in figure~\ref{fig_sigl} (right) results in
consistent values of \asmz\ and \anulltwo\ with large errors. The
presently available measurements do not allow quantitative studies of
the power correction.

\section{ Summary }

We have shown experimental tests of power corrections in the model of
Dokshitzer, Marchesini and Webber using data from \epem\ annihilation
experiments. The DMW model together with perturbative QCD describes
well distributions and mean values of selected event shape
observables. The results for \anull\ from the individual observables
are generally consistent within the total uncertainties. 
A measurement of the QCD colour factors using the same
QCD predictions combined with power corrections gave results
consistent with QCD based on SU(3). A new measurement of the
longitudinal cross section $\sigl(36.6\;\mathrm{GeV})$ was presented
and compared with a QCD prediction including power corrections.

\end{document}